\documentclass[twocolumn]{aastex63}
\usepackage{multirow}
\usepackage{amsmath}
\usepackage{booktabs}

\usepackage{xcolor}

\newcommand\lsim{\mathrel{\rlap{\lower4pt\hbox{\hskip1pt$\sim$}}
\raise1pt\hbox{$<$}}}
\accepted{ApJ}

\shorttitle{SIGOs through the Universe}
\shortauthors{Lake et al.}
\graphicspath{{./}{figures/}}

\begin{document}

\title{ The Supersonic Project: SIGOs, a Proposed Progenitor to Globular Clusters, and their Connections to Gravitational Wave Anisotropies}


\correspondingauthor{William Lake}
\email{wlake@astro.ucla.edu}
\author[0000-0002-4227-7919]{William Lake}
\affil{Department of Physics and Astronomy, UCLA, Los Angeles, CA 90095}
\affil{Mani L. Bhaumik Institute for Theoretical Physics, Department of Physics and Astronomy, UCLA, Los Angeles, CA 90095, USA\\}

\author[0000-0002-9802-9279]{Smadar Naoz}
\affil{Department of Physics and Astronomy, UCLA, Los Angeles, CA 90095}
\affil{Mani L. Bhaumik Institute for Theoretical Physics, Department of Physics and Astronomy, UCLA, Los Angeles, CA 90095, USA\\}

\author[0000-0003-4962-5768]{Yeou S. Chiou}
\affil{Department of Physics and Astronomy, UCLA, Los Angeles, CA 90095}
\affil{Mani L. Bhaumik Institute for Theoretical Physics, Department of Physics and Astronomy, UCLA, Los Angeles, CA 90095, USA\\}

\author[0000-0001-5817-5944]{Blakesley Burkhart}
\affiliation{Department of Physics and Astronomy, Rutgers, The State University of New Jersey, 136 Frelinghuysen Rd, Piscataway, NJ 08854, USA \\}
\affiliation{Center for Computational Astrophysics, Flatiron Institute, 162 Fifth Avenue, New York, NY 10010, USA \\}

\author[0000-0003-3816-7028]{Federico Marinacci}
\affiliation{Department of Physics \& Astronomy ``Augusto Righi", University of Bologna, via Gobetti 93/2, 40129 Bologna, Italy\\}

\author[0000-0001-8593-7692]{Mark Vogelsberger}
\affil{Department of Physics and Kavli Institute for Astrophysics and Space Research, Massachusetts Institute of Technology, Cambridge, MA 02139, USA\\}

\author[0000-0002-4086-3180]{Kyle Kremer}
\affiliation{TAPIR, California Institute of Technology, Pasadena, CA 91125, USA}
\affiliation{The Observatories of the Carnegie Institution for Science, Pasadena, CA 91101, USA}



\begin{abstract}
Supersonically Induced Gas Objects (SIGOs), are structures with little to no dark matter component predicted to exist in regions of the Universe with large relative velocities between baryons and dark matter at the time of recombination. They have been suggested to be the progenitors of present-day globular clusters. Using simulations, SIGOs have been studied on small scales (around 2 Mpc), where these relative velocities are coherent. However, it is challenging to study SIGOs using simulations on large scales due to the varying relative velocities at scales larger than a few Mpc. Here, we study SIGO abundances semi-analytically: using perturbation theory, we predict the number density of SIGOs analytically, and compare these results to small-box numerical simulations. We use the agreement between the numerical and analytic calculations to extrapolate the large-scale variation of SIGO abundances over different stream velocities. As a result, we predict similar large-scale variations of objects with high gas densities before reionization that could possibly be observed by JWST. If indeed SIGOs are progenitors of globular clusters, then we expect a similar variation of globular cluster abundances over large scales. Significantly, we find that the expected number density of SIGOs is consistent with observed globular cluster number densities. 
As a proof-of-concept, and because globular clusters were proposed to be natural formation sites for gravitational wave sources from binary black hole (BBH) mergers, we show that SIGOs should imprint an anisotropy on the gravitational wave signal on the sky, consistent with SIGOs' distribution. 
\end{abstract}


\section{Introduction}




Globular clusters (GCs) are very old \citep[$\sim$~13 Gyr, e.g.,][]{Trenti+15} structures with masses between $\sim 10^5 - 10^6 M_\odot$ \citep[e.g.,][]{Elmegreen+97,Fall+01,McLaughlin+08,Elmegreen+10}. 
Their high stellar densities and low metallicities make them a promising nurturing ground for gravitational wave sources via few body dynamics \citep[e.g.,][]{Zwart_2000,Wen03,OLeary_2006,Rodriguez+15,Rodriguez+16,Chatterjee+17,Kremer+20,Rodriguez+21}.

Significantly, observations suggest that GCs contain little to no dark matter \citep[e.g.,][]{Heggie+96,Bradford+11,Conroy+11,Ibata+13}. 
These observations pose a challenge to the formation of these objects in the context of hierarchical structure formation.
Accordingly, different GC formation scenarios exist  in the literature. One popular mechanism is that GCs formed as a byproduct of active star formation in galaxy discs \citep[e.g.,][]{Elmegreen+10,Shapiro2010,Kruijssen2015}, for example as a result of strong shocks when gas is compressed during galaxy mergers, as first proposed by \citet[]{Gunn80}. The discovery of many massive
young star clusters in the interacting Antennae system
\citep[e.g.,][]{Whitmore+95,Whitmore+99} supports this idea.  Furthermore, this scenario has also been incorporated into cosmological hierarchical structure formation models \citep[e.g.][]{Harris+94,Ashman+92,Kravtsov+05,Muratov+10}. However, this paradigm is challenged by observations of nuclear star clusters that resemble GCs (e.g., in total mass and core/half-light radii),  which imply that some GC-like structures may form inside dark matter (DM) halos and thus may have a DM halo origin
\citep[see for example][]{Boker+04,Walcher+05,Walcher+06,Brown+14}. 

Another popular theory is that GCs initially formed inside dark matter halos \citep[as suggested by][]{Peebles84}, but that these halos
were later stripped by the tidal field of their host galaxies, leaving
the central parts deficient of dark matter \citep[e.g.][]{Bromm+02,Mashchenko+05II,Saitoh+06,Bekki+12}.  However, some GCs are observed
to have stellar tidal tails, which is difficult to explain in the context of this scenario. If the
objects have extended dark matter halos, the halos should have shielded them from forming tidal tails
\citep[e.g.,][]{Grillmair+95,Moore96,Odenkirchen+03,Mashchenko+05II}.


Recently, \citet{naoznarayan14} proposed a formation pathway for GCs that relies on the relative motion between baryons and DM at the time of recombination, known as the stream velocity \citep{TH,tseliakhovich11}.
In the standard model of structure formation, 
due to the baryon-photon coupling, dark matter began to collapse to form overdensities far more efficiently than baryons. By the time of recombination, when baryons decoupled from photons, baryon overdensities were about $5$ orders of magnitude smaller than dark matter overdensities \citep[e.g.][]{NB}. This meant that the existing dark matter overdensities dominated the dynamics of baryon overdensity formation. 
\citet{TH} showed that in addition to the difference in amplitude between baryonic and DM overdensities, there was a significant difference in their velocities in the period following recombination. As the baryons cooled, the typical relative velocity between dark matter and baryons (about 30 km s$^{-1}$) became supersonic. They also showed that this relative velocity was coherent on scales of $\sim 2-3$~Mpc, allowing it to be modelled as a stream velocity on these scales.

This stream velocity suppresses formation of the earliest baryonic structures, such as minihalos, and therefore has an effect on early star formation and on the temperature of the early Universe. This has been explored in a variety of studies, having such diverse impacts as creating temperature-induced fluctuations in the cosmological 21 cm line \citep[e.g.,][]{dalal10,Visbal+12,McQuinn+12,Cain+20}, enhancing primordial black hole formation \citep[e.g.,][]{tanaka13,tanaka14,latif14,Hirano+17,schauer17}, and even creating primordial magnetic fields through temperature fluctuation-induced vorticity \citep{naoznarayan13}. Studies also show that this stream velocity has major impacts on the number densities of halos \citep{asaba16,tanaka13,tanaka14,bovy13,oleary12,naoz12,fialkov2012,TH,Maio}, as well as the overall gas fraction in halos \citep{asaba16,richardson13,Maio,oleary12,Naoz+13,greif11,fialkov2012,naoz12,tseliakhovich11,dalal10} and the size of halos able to retain gas at each redshift \citep{Naoz+13}. In addition, the stream velocity impacts the gas density and temperature profiles \citep{richardson13,oleary12,fialkov2012,greif11,Maio,liu11,Druschke+20}, and the halo mass threshold at which star formation occurs \citep{bovy13,oleary12,fialkov2012,greif11,Maio,liu11, schauer+19}.

The aforementioned proposal by \citet[]{naoznarayan14} suggested that this stream velocity effect could lead to a possible  formation mechanism for globular clusters. They found that a stream velocity of sufficient magnitude between a dark matter and baryonic overdensity could create a spatial offset between the collapsing baryonic overdensity and its parent dark matter halo. In certain instances, this effect is large enough to cause the baryonic overdensity to collapse outside the parent halo's virial radius, allowing it to be separated from the parent halo's gravitational influence entirely. This would create a baryonic clump depleted of dark matter in a similar mass range to present-day globular clusters. In addition, such a baryon clump would likely have a low metallicity attributable to its early formation, possibly consistent with that of the low-metallicity population of GCs.

These objects, known as Supersonically Induced Gas Objects (SIGOs), have since been found in follow-up simulations \citep{popa,chiou18,chiou+19,Chiou+21}. 
However, there are still many outstanding questions about these objects. Notably,  their large-scale abundance distribution has not been studied yet. This large-scale abundance is expected to be correlated with the magnitude of the stream velocity \citep{popa}. SIGO abundances determine their possible global effect on reionization as well as the distribution of the very first star clusters and possibly GCs.

The question of the connection between SIGO abundances and GC abundances has particular relevance given the recent detections of gravitational wave (GW) emission from merging stellar-mass black hole (BH) binaries by LIGO-Virgo that have expanded our ability to sense the Universe \citep[e.g.,][]{Abbott+16,Abbott+17}. It remains challenging to explain the formation channels of these sources, but recent studies have emphasized the significant contribution of dynamical formation channels in dense stellar environments to the overall population of GW signals \citep[e.g.,][]{Zwart_2000,Wen03,O'Leary+06,O'Leary+09,O'Leary+16,Kocsis+12,Antonini+15,Rodriguez+16PhRvD,Chatterjee+17,Stone+16,Hoang+18,Stephan+19,Kremer+20,Wang+20}. Accordingly, GCs have been suggested as a primary source of black-hole binary (BBH) mergers \citep[e.g.,][]{Rodriguez+21}. Should this be the case, and should SIGOs indeed be connected to GCs, then SIGO abundances should be connected to the abundance of BBH mergers.

In this paper, we study the large-scale abundances of SIGOs using a combination of analytical and numerical methods. This is a challenging task due the following reasons:
\begin{itemize}
    \item The stream velocity is constant only on scales of a few Mpc \citep[e.g.,][]{TH}. Thus, the implementation of the initial conditions in numerical simulations can be done self-consistently only on small box simulations \citep[e.g.,][]{naoz12,Naoz+13,McQuinn+12,oleary12,Stacy+10,schauer+19,chiou18,chiou+19,Chiou+21}. In these small box simulations, the stream velocity is implemented as a uniform boost along one axis. 
    \item Even at the event of successfully implementing initial conditions that allow for the stream velocity to change coherently over large scales ($>>$~few Mpc), the simulation will still need to resolve objects at the order of $10^4$~M$_\odot$ with at least around $100$ particles, requiring unrealistic numerical resources.  
\end{itemize}
We therefore take a combined approach, utilizing analytical and numerical tools. We use a series of small-box {\tt AREPO} runs (side length $2$~Mpc) with varying stream velocity magnitudes and compare them to analytical calculations. 
Using simulation results to derive an abundance normalization factor, we create a fully analytic model of the spatial variation of SIGO abundances.
If SIGOs are indeed linked to GCs, they can host gravitational wave sources, which allows us to hypothesize a spatial variation in GC and GW abundances related to that of SIGOs.

For this work, we have assumed a $\Lambda$CDM cosmology with $\Omega_{\rm \Lambda} = 0.73$, $\Omega_{\rm M} = 0.27$, $\Omega_{\rm B} = 0.044$, $\sigma_8  = 1.7$, and $h = 0.71$.



This paper is organized as follows: we first provide an overview of our simulations in Section \ref{ssec:sim}. We then discuss our analytic model in Section \ref{ssec:Analytic}. We provide a comparison between the simulation and model results in Section \ref{sec:comparison}, as well as connecting our model results to the real-world abundance of GCs. We consider the implications of these results to gravitational wave abundances in Section \ref{sec:GW}. We discuss our model results in Section \ref{sec:discussion}. Finally, we show how we normalized our analytic model to simulations in Appendix \ref{sec:appendix} and provide an analytic approximation to our model in Appendix \ref{Appendix-B}.



\section{Methods}\label{sec:Meth}

We use a combination of analytical and numerical methods described below to analyze the large-scale SIGOs number density. 

\subsection{Simulations}\label{ssec:sim}


We present three simulations with the moving-mesh code {\tt AREPO} \citep{springel10} in a $2$~Mpc box\footnote{Note that the simulated abundances of SIGOs at the relevant masses
were shown to converge for small (few Mpc) boxes \citep[e.g.,][]{popa}.} with $512^3$ DM particles of mass $M_{\rm DM}=1.9\times10^3$~M$_\odot$ and $512^3$ Voronoi mesh cells with $M_{\rm b}=360$~M$_\odot$, evolved from $z=200$ to $z=20$. These runs had stream velocities of $v_{\rm bc}=1 \sigma_{\rm vbc}$, $2 \sigma_{\rm vbc}$, and $3 \sigma_{\rm vbc}$ where $\sigma_{\rm vbc}$ is the rms value of the stream velocity--the relative velocity of the gas component with respect to the dark matter component.
$\sigma_{\rm vbc} = 5.9$~km sec$^{-1}$ at $z=200$. We note that these runs do not include radiative cooling. Cooling does not significantly change the physical properties of SIGOs, and only moderately affects the classical objects (i.e., DM halos with gas), as shown in \citet{Chiou+21}. 

The initial conditions for our cosmological simulations were generated using transfer functions calculated using a modified {\tt CMBFAST} code \citep{seljak96} that takes into account the first-order correction of scale-dependent temperature fluctuations \citep{NB}.  These transfer functions also include second-order corrections to the equations presented in \cite{TH} that describe the evolution of the stream velocity. There are two transfer functions, one for the baryons and one for the dark matter, as it was pointed out that the gas fraction evolution strongly depends on the baryons' initial conditions \citep[e.g.,][]{Naoz+09,naoz11,Naoz+13,park+20}. The stream velocity was implemented in the initial conditions as a uniform boost to the gas in the x-direction, as in \cite{popa}. Initial conditions were generated at $z=200$.

For this paper, we use the object classifications described in \citet[]{chiou18}. The first step in our identification of SIGOs is to identify dark matter-primary objects (dark matter halos) using a Friends-of-Friends (FOF) algorithm with a linking length that is $20\%$ of the mean particle separation on the DM component of the simulation output\footnote{This linking length was shown to give converging values of object abundances by \citet[]{naoz11}}, about 780 comoving pc. This algorithm identifies the location of the DM halos in the simulation box. It also calculates the virial radius for each halo, assuming sphericity for simplicity \citep[although DM halos show distinct triaxiality e.g.,][]{Sheth+01,Lithwick+11,Vogelsberger+11,Schneider+12,Vogelsberger+20}. Next, we find gas-primary objects using the same FOF algorithm run only on the gas component of the simulation output. We require that gas primary objects contain at least 32 particles to be considered a SIGO \citep[][]{Chiou+21}. Because these objects tend to be more attenuated, each gas-primary object is fit to an ellipsoid, by identifying an ellipsoidal surface that encloses every particle in the gas object \citep[][]{popa}. We then tighten these ellipsoids by shrinking their axes by $5\%$ until either $20\%$ of their particles have been removed, or until the ratio of the axes lengths of the tightened ellipsoid to that of the original ellipsoid is greater than the ratio of the number of gas cells contained in each, as in \citet[]{popa}. Because many of these gas-primary objects are actually just the gas component of the previously mentioned DM halos, SIGOs are then defined as gas-primary objects which have a gas fraction above $40\%$\footnote{Note that 40\% here represents a somewhat arbitrary compromise between doubling the cosmic baryon fraction and the estimated baryon content of GCs, 50\% or more.}, and are outside the virial radius of the nearest dark matter halo.

\subsection{Analytic Model}\label{ssec:Analytic}

Our analytic model, in contrast to our simulations, ran on a large-scale box ($\sim$ $1365$~Mpc on a side), composed of grid cells that were $3$~Mpc on a side. Within each grid cell, as in the simulations, the relevant scales are small enough that $v_{\rm bc}$ is approximately constant. We assigned a value of $v_{\rm bc}$ for each cell using an algorithm for generating Maxwell distributed random fields given a power spectrum of their spatial fluctuations \citep{Brown_2013} which we calculated using a modified version of {\tt CMBFAST} \citep{seljak96}, that includes the spatial perturbations of the baryon sound speeds, as outlined in \citet[]{NB}. 


Stream velocities follow a Maxwell distribution with scale parameter $\sigma = \sigma_{\rm vbc}/\sqrt{3}$, and a known power spectrum given by the output of {\tt CMBFAST} described above. Using the spectral distortion method outlined in \citet[]{Brown_2013}, we generated a Maxwell-distributed random field of velocities in a grid of $456 \times 456 \times 456$ cells, with each cell being 3 Mpc on a side (small enough such that the stream velocity within each cell is coherent). This grid was generated with the computed power spectrum through the following recursive steps: 

\begin{enumerate} 
\item We generated a Gaussian random field using the computed power spectrum of stream velocity fluctuations $P_{\rm vbc}$ as the input power spectrum $P_{\rm I}$. 

\item We then transformed this Gaussian random field to a Maxwell distributed random field using a quantile transform, and calculated the output power spectrum $P_{\rm F}$ of that field. 

\item If this power spectrum output is consistent with the target output, we accepted this Maxwell-distributed field as our velocity grid. Otherwise, we set our input power spectrum

\begin{equation}\label{Eq:InputPowerSpectrum}
P_{\rm I, new} = \frac{P_{\rm F}(k)}{P_{\rm I}(k)} \times P_{\rm vbc}(k) \ ,
\end{equation}

and returned to the first step using this new input power spectrum. 

\end{enumerate}
This yields a grid of cells with constant stream velocity to be used in the density evolution equations that follow.

For completeness, we provide the full set of differential equations of the perturbation theory. 
We solve the differential equations for the dimensionless overdensities of both the dark matter, $\delta_{\rm dm}$, and the baryons, $\delta_{\rm b}$ in the presence of the relative velocity between dark matter and baryons in small regions within which the velocity is coherent \citep[few Mpc, e.g.][]{TH,tseliakhovich11,Naoz+13}. These can be expressed by the following set of coupled equations:
\begin{align}
&\begin{aligned}
\begin{split}\label{eq:1}
\ddot{\delta}_{\rm dm} +& 2 H \dot{\delta}_{\rm dm} - f_{\rm dm} \frac{2i}{a} \mathbf{v}_{\rm bc} \cdot \mathbf{k} \dot{\delta}_{\rm dm} =  \\ & \frac{3}{2} H_0^2 \frac{\Omega_{\rm m}}{a^3} (f_{\rm b} \delta_{\rm b} + f_{\rm dm}\delta_{\rm dm}) + (\frac{\mathbf{v_{\rm bc}}\cdot\mathbf{k}}{a})^2 \delta_{\rm dm} \ ,
\end{split}
\end{aligned}
\end{align}
\begin{align}
&\begin{aligned}
\begin{split}\label{eq:2}
\ddot{\delta}_{\rm b} + 2 H \dot{\delta}_{\rm b}  = \frac{3}{2} H_0^2 \frac{\Omega_{\rm m}}{a^3}& (f_{\rm b} \delta_{\rm b} + f_{\rm dm}\delta_{\rm dm}) \\ & - \frac{k^2}{a^2} \frac{k_{\rm b} \bar{T}}{\mu} (\delta_{\rm b} + \delta_{\rm T}) \ ,
\end{split}
\end{aligned}
\end{align} 
where ${\bf k}$ is the comoving wavenumber vector, $a$ is the scale factor, $\mu$ is the mean molecular weight, $\delta_{T_\gamma}$ is the photon temperature fluctuations, $\bar{T}$ and $f_{\rm b}$ and $f_{\rm c}$ are the baryon and DM fractions, respectively.

 We also include the baryons' temperature fluctuations $\delta_{\rm T}$ which include scale-dependent temperature time evolution \citep[according to][]{NB} in our calculations. These evolve according to
\begin{eqnarray}\label{Eq. 3}
\frac{{\delta}_{\rm T}}{dt} &=&  \frac{2}{3} \frac{\delta_{\rm b}}{dt}  \\ &+&  \frac{x_{\rm e}(t)}{t_{\rm \gamma}} a^{-4} \left\{\delta_{\rm \gamma} \left(\frac{\bar{T}_{\rm \gamma}}{\bar{T}} - 1 \right) + \frac{\bar{T}_{\rm \gamma}}{\bar{T}} \left(\delta_{T_{\rm \gamma}} - \delta_{\rm T} \right) \right\} \  . \nonumber
\end{eqnarray}
As was discussed in \citet{naoz12} and \citet{naoznarayan14}, the stream velocity introduces a phase shift between baryon and DM overdensities. This phase shift creates a spatial separation between baryonic overdensities and their parent DM overdensities. Because increasing stream velocities create increasing phase shifts (which in turn create increasing spatial separations between the overdensities), sufficiently high stream velocities can cause baryonic clumps to collapse outside of the virial radii of their parent DM overdensities. This allows them to survive as independent, DM-depleted objects. Simulations suggest that these objects could potentially evolve into present-day globular clusters \citep{chiou+19}.


\begin{figure*}[t]
\includegraphics[width=1\textwidth]{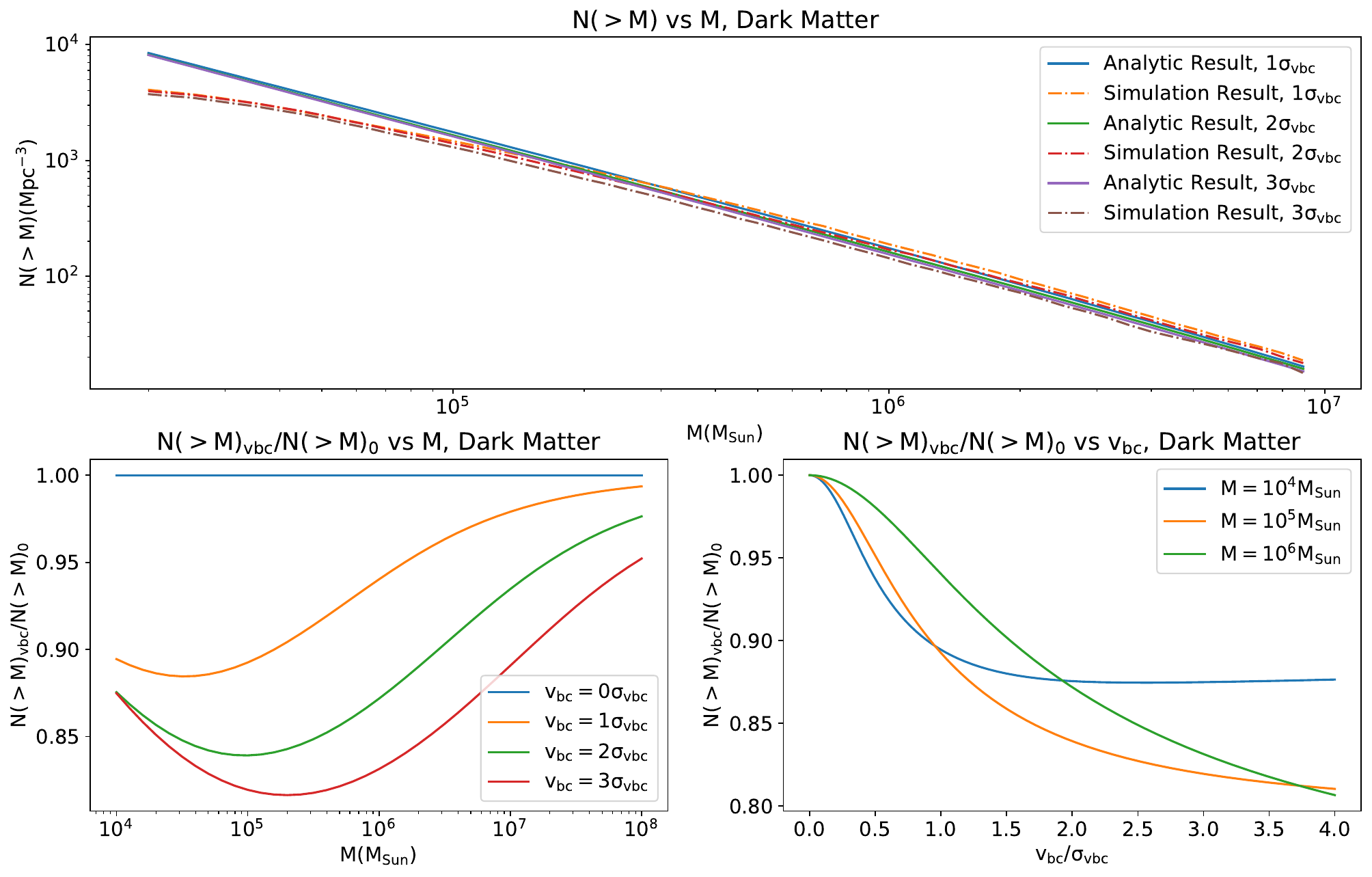}
\caption{{\bf The effect of the stream velocity on the ``classical'' dark matter halos.} In the top panel, we consider the number abundance $N(>M)_{\rm Halo, vbc}$ as a function of the DM halo mass, and compare the analytical calculation (solid lines) and simulation results (dot-dashed lines). In the bottom panels, we consider the analytical reduction fraction of the number of halos due to the stream velocity (i.e., $N(>M)_{\rm Halo, vbc}/N(>M)_{Halo, 0}$, where subscript ``$0$'' indicates $v_{\rm bc}=0$). We present this fraction as a function of DM halo mass, for different stream velocity values (bottom left panel) and as a function of the stream velocity for different halo masses (bottom right panel). Results are for $z=20$.  }\label{fig:MtotComp}
\end{figure*}

We adopt the generalized Press-Schechter formalism \citep{Press+74} to allow for non-spherical halos. This model, based on Gaussian random fields and including linear growth, allows us to calculate abundances of objects at different masses. The formalism depends on two functions $\sigma(M,z)$ and $\delta_c $. In this case, $ \sigma^2 (M,z) $ is the variance, calculated from the power spectrum, as a function of halo mass at a given redshift, and $ \delta_c  $ is the critical collapse overdensity\footnote{We note that in general  $ \delta_c $ is a function of the redshift \citep[e.g.][]{Naoz+06,Naoz+07} because the baryons have smoother initial conditions. However, since we normalize our abundances according to the simulations, we neglect the redshift contribution.}.
The comoving number density of halos of mass M at redshift z in this model is given by 
\begin{equation} 
\frac{dn}{dM}  = \frac{\rho_0}{M}f_{ST}\left|\frac{dS}{dM}\right| \ , \end{equation} where we have used the \citet{Sheth+01} mass function that both fits simulations, and includes non-spherical effects on the collapse. The function $f_{ST}$ is the fraction of mass in halos of mass M:
\begin{eqnarray}
 f_{ST}(\delta_c, S) &=&  A' \frac{\nu}{S}\sqrt{\frac{a'}{2\pi}}\left[1+\frac{1}{(a'\nu^2)^{q'}}\right]\exp\left(\frac{-a'\nu^2}{2}\right) \  \\
{\rm and} \quad \nu&=&\frac{\delta_c}{\sigma}=\frac{\delta_c}{\sqrt{S}} \ . \nonumber\end{eqnarray} 
We use best-fit parameters $A'=0.75$ and $q'=0.3$ \citep{Sheth+02}. By evolving the power spectrum analytically, we can use this model to effectively predict halo abundances, as shown in Section \ref{sec:comparison}. 

It not straightforward to extrapolate the DM halo Press-Schechter formalism to SIGOs because these objects are non-spherical. Significantly, unlike a DM overdensity that grows due to its own gravity, SIGOs by themselves do not have enough material to grow independently \citep{peebles}, and instead are still coupled to the DM potential wells \citep{naoznarayan14}. In the presence of the stream velocity the gas does not accumulate over the DM overdensities \citep[e.g.,][]{naoz12,popa} and some of it results in the formation of SIGOs. Thus, we postulate that the SIGOs overdensity may be related to the DM underdensity, and could obey a simple relation such as:
\begin{equation}\label{eq:NSIGOs}
    \frac{dN(v_{\rm bc})}{dM}\bigg |_{\rm SIGO} \propto  \left( \frac{dN(v_{\rm bc}=0)}{dM}\bigg |_{\rm Halo}-\frac{dN(v_{\rm bc})}{dM}\bigg |_{\rm Halo} \right) \ ,
\end{equation}
where the proportionality here is aimed to emphasize that the non-linear effects result in a normalization factor (see Appendix \ref{sec:appendix} for details). In other words, some of the gas that does not fall onto the DM potential wells does not become SIGOs.  Motivated by simulations, we find a simple normalization power law in v$_{\rm bc}$ (equation [\ref{eq:NSIGOs2}]).

By evolving the power spectrum analytically, we can determine an analytical SIGO abundance, and thereby determine properties of their distribution on large scales (i.e., the sky). We can then compare this predicted spatial variation of SIGO abundances to observations of globular cluster abundances to test the hypothesis that these objects are their dominant formation mechanism.

\section{Comparison between analytical and numerical calculations}\label{sec:comparison}
\subsection{Dark Matter Halos}

In Figure \ref{fig:MtotComp} we show the agreement between the analytic model based on the generalized Press-Schechter formalism and simulation results for DM halos, for various values of the stream velocity effect at $z=20$. As depicted, the simulations and the analytical calculations are consistent for $M\approx 10^5-10^7 M_\odot$, the region in which the simulation results are expected to be less sensitive to resolution effects (requiring a minimum of $100$ DM particles per halo).
Thus, because of the limited resolution of the simulation, we observe fewer small-mass halos in the simulation than our analytic model would predict, as expected.

The bottom panels of Figure \ref{fig:MtotComp}, present only analytical calculations. We show the fraction of DM number density $N(>M)_{\rm Halo}$ with the stream velocity compared to the number density without the stream velocity. In other words:
\begin{equation}\label{eq:fDM}
    f_{\rm DM}=\frac{N(>M)_{\rm Halo, vbc}}{N(>M)_{{\rm Halo},0}} \ ,
\end{equation}
 where the subscript ``$0$'' means $v_{\rm bc}=0$ and the cumulative comoving number density of DM haloes $N(>M)$ is given by
\begin{equation}\label{eq:NofM}
    N(>M)_{\rm Halo} =\int^{\infty}_{{\rm M}} \frac{dN_{\rm Halo}}{dM} dM  \ .
\end{equation}

\begin{figure}[t]
\includegraphics[width=.475 \textwidth]{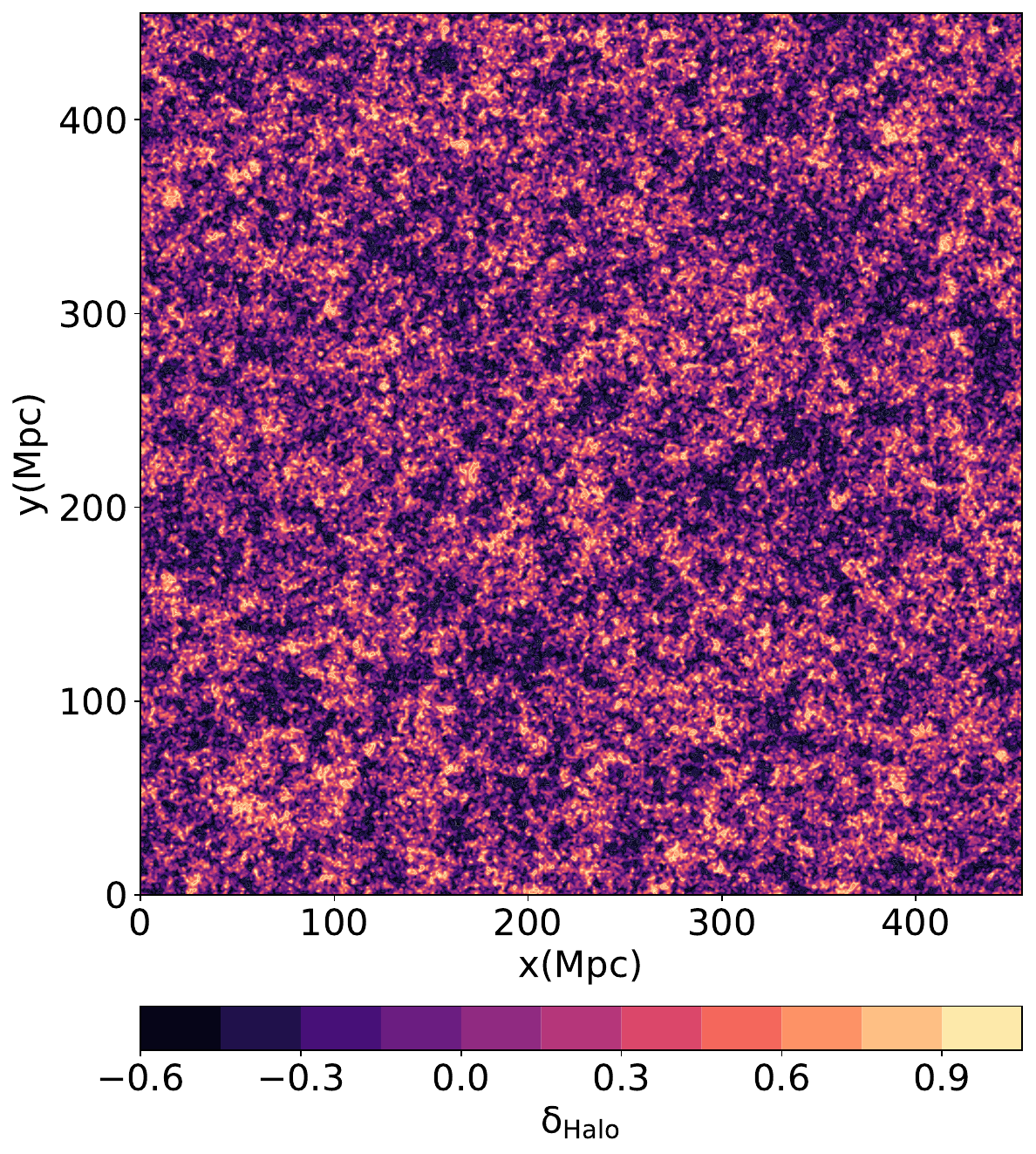}
\caption{Map of DM halo density contrast at $z=20$, showing a $456$ Mpc $\times$ $456$ Mpc $\times$ $1$ Mpc box (a subset of our model box, to enhance the visibility of structures).}\label{fig:DMMap}
\end{figure}

The bottom left panel of Figure \ref{fig:MtotComp} shows Equation (\ref{eq:fDM}) as a function of the DM halo mass for different stream velocity effects. The bottom right panel shows Equation (\ref{eq:fDM}) as a function of $v_{\rm bc}$, for different DM halo masses.  As depicted, the higher stream velocities reduce the abundance of dark matter halos, particularly those of mass $\sim 10^5-10^6 M_\odot$, with the total reduction in abundance with stream velocities on the order of $\sigma_{\rm vbc}$ being on the order of tens of percent.

\begin{figure*}[t]
\includegraphics[width=1\textwidth]{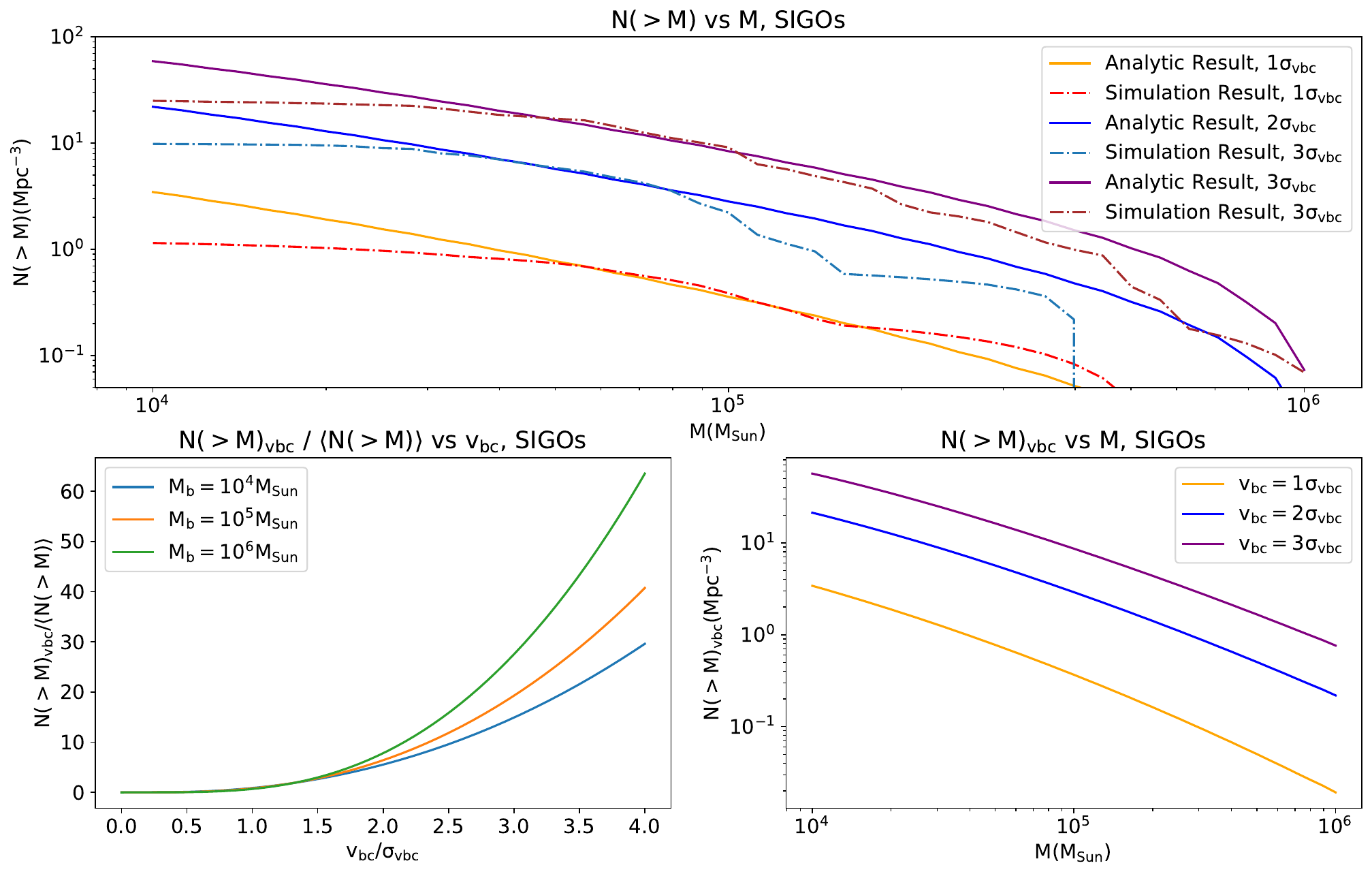}
\caption{{\bf The effect of the stream velocity on SIGO abundances.} In the top panel, we consider the number abundance $N(>M)_{\rm SIGO, vbc}$ as a function of SIGO mass, and compare the analytical calculation (solid lines) and simulation results (dot-dashed lines). See text for an explanation of how SIGO abundances are estimated in the analytical calculations and how they are defined in the simulations, and see Equation (\ref{eq:meanSIGO}) for a definition of $\left<N(>M)_{\rm SIGO}\right>$. In the bottom left panel, we consider the number density of SIGOs as a function of the stream velocity magnitude for different SIGO masses, based on analytical calculations. In the bottom right panel, we depict the number density of SIGOs as a function of mass for different stream velocities, based on analytical calculations, without integrating from a maximum mass. Results are for $z=20$.}\label{fig:MSIGOComp}
\end{figure*}

In Figure \ref{fig:DMMap} we depict the DM halo fluctuations due to the stream velocity effects on a large scale. In particular, we show  
\begin{equation}\label{eq:delta}
    \delta_{\rm Halo} = \frac{N(>M)_{\rm Halo, vbc} - \left<N(>M)_{\rm Halo}\right>}{\left<N(>M)_{\rm Halo}\right>} \ ,
\end{equation}
with $N(>M)_{\rm Halo}$ as defined in Equation (\ref{eq:NofM}), and using the velocity field generated with the method described in Section \ref{ssec:Analytic}.




\begin{figure}[t]
\includegraphics[width=0.475\textwidth]{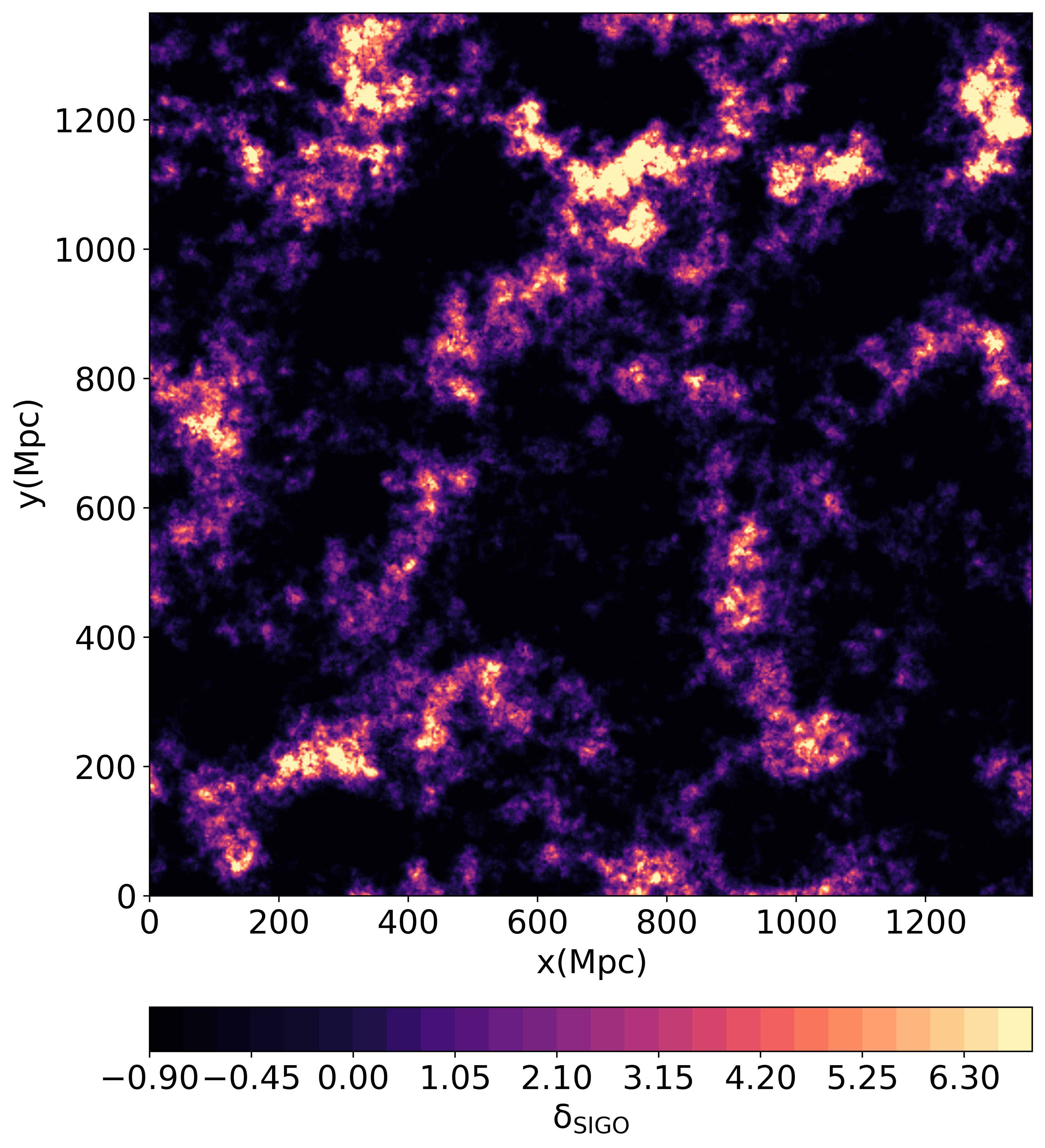}
\caption{Map of SIGO density contrast at z=20. Specifically, we plot Eq.~(\ref{eq:deltaSIGO}), for a minimum mass of $10^5$ $M_\odot$. Here the color scale was capped at $\delta_{\rm SIGO} = 7.0$ to enhance visibility of smaller fluctuations; the true maximum in this plot is $\delta_{\rm SIGO} \approx 30$(which represents a rare fluctuation).}\label{fig:SIGOMap}
\end{figure}

\subsection{SIGOs}\label{sec:SIGOs}

Using the tightly fitted ellipsoid method to find the SIGOs (see Section \ref{sec:Meth}) we find $8, 75$ and $188$ SIGOs for the $1,2$ and $3\sigma_{\rm vbc}$ simulation runs. From this simulation data, we construct $N(>M)_{\rm SIGO, sim}$.
As discussed in Section \ref{ssec:Analytic}, we also calculate the abundance of SIGOs as a function $\sigma_{\rm vbc}$
analytically using Equation (\ref{eq:NSIGOs}). We normalize the analytical results to the simulation results as described in appendix \ref{sec:appendix}.
The comparison yields a simple functional form for SIGO abundances as function of mass and stream velocity, i.e.,
\begin{align}
&\begin{aligned}
\begin{split}\label{eq:NSIGOFunc}
    \frac{dN(u_{\rm vbc})}{dM}\bigg |_{\rm SIGO} \approx & \left(\frac{ u_{\rm vbc}}{20.76}\right)^{2.43} \times \\ & \left( \frac{dN(u_{\rm vbc}=0)}{dM}\bigg |_{\rm Halo}-\frac{dN(u_{\rm vbc})}{dM}\bigg |_{\rm Halo} \right) \ ,
\end{split}
\end{aligned}
\end{align}
with 
$u_{\rm vbc}=v_{\rm bc}/\sigma_{\rm vbc}$. At the time of recombination, for example, $\sigma_{\rm vbc}=30$~km~sec$^{-1}$, which corresponds to $u_{\rm vbc} = 1$.  

\begin{figure*}[htp]
\includegraphics[width=\textwidth]{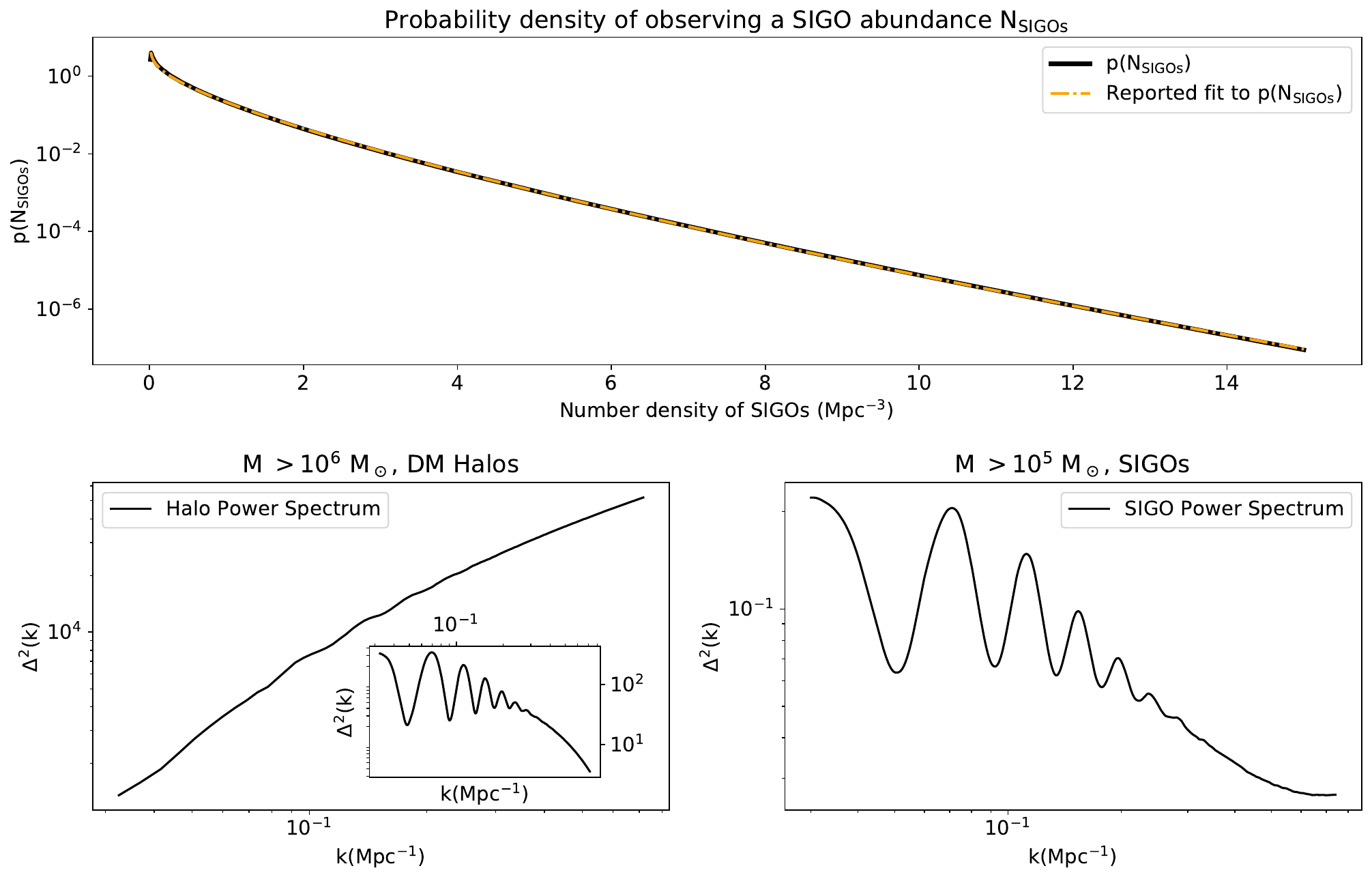}
\caption{\textbf{The distribution of SIGOs in the Universe.} In the top panel, we show the probability density $p(N_{\rm SIGO})$ of observing a given abundance of SIGOs with M $> 10^5 {\rm M_\odot}$ in a randomly selected region of the Universe with an unknown (but constant) stream velocity. We also show a fit to this probability density function, given by Equation (\ref{eq:pSIGOApprox}). In the bottom left panel, we present the power spectrum of the distribution of ${\rm M}_{\rm Halo}>10^6$ ${\rm M_\odot}$ halos in our analytic model. Displayed is $\Delta^2 (k) \equiv k^3/(2\pi^2) P(k)$, a non-dimensional quantity describing the variance in N($>$M)$_{\rm Halo}$ per $\ln k$. Inset in the bottom left panel, we also show the portion of the same power spectrum that is solely the result of the stream velocity, and not an effect of large scale density fluctuations. In the bottom right panel, we show the power spectrum of the abundances of SIGOs with M~$> 10^5 {\rm M_\odot}$ in our analytic model, also given as $\Delta^2 (k)$. Note the different y-scale in the two bottom panels.}\label{fig:CombinedPSPlot}

\end{figure*}

Figure \ref{fig:MSIGOComp}, top panel, shows the  comparison between the SIGOs model and simulation results (as in Figure \ref{fig:MtotComp}). In the analytical model, we have integrated the number density $dN/dM$ (see Equation [\ref{eq:NSIGOFunc}]) only from a defined cutoff mass (${\rm M}_{\rm cutoff} =  M_{\rm SIGO, max} = 1.1 \times 10^6 M_\odot$) 
for the purpose of comparing to simulations. This cutoff mass is motivated by \citet[]{naoznarayan14}, who found that SIGOs have an upper mass limit of around a few $\times 10^6$ ${\rm M_\odot}$, above which they are incapable of escaping their parent DM halo. Because of this, we select ${\rm M_{cutoff}}$ as the mass of the largest SIGO observed in any of our simulations.
In particular, the cumulative number density  can be expressed  as:
\begin{equation}\label{eq:NofMSIGO}
    N(>M_{\rm min}) =\int^{M_{\rm SIGO, max}}_{{\rm M}_{\rm min}} \left(\frac{dN}{dM}\right)_{\rm SIGO} dM  \ ,
\end{equation}
where $dN/dM$ is defined in Eq.~(\ref{eq:NSIGOFunc}). We show the analytical calculations based on this equation in Figure \ref{fig:MSIGOComp}, solid lines, in all panels. From top to bottom, we consider $1,2,$ and $3~\sigma_{\rm vbc}$ effects, and compare these effects to the cumulative SIGO abundance estimated from our simulation boxes, shown as dashed lines. Note the consistency between the analytical and simulation SIGOs abundances at the range of few$\times 10^4$-few$\times 10^5$~M$_\odot$.

At small masses ($M\approx10^4$~M$_\odot)$, the limited resolution of the simulation yields lower abundances.
At large masses, Poisson fluctuations increase the uncertainty of our simulation results, again creating an apparent disagreement between our simulation results and analytic model\footnote{For example, the largest such fluctuation,for  M$\sim$few$\times 10^5 {\rm M_\odot}$, is about a $3 \sigma$ deviation in the $2 \sigma_{\rm vbc}$ data. This represents a $1\%$ fluctuation, which is consistent with the largest fluctuations due to Poisson statistics we would expect in our data set.}. Given these caveats and the agreement over the relevant range we are confident that the analytical model provides a reasonable approach to estimate SIGO abundances.

Therefore, the bottom panels of Figure \ref{fig:MSIGOComp} show analytical calculations only, analogous to those of Figure \ref{fig:MtotComp}. The bottom right panel shows our analytical results for comoving SIGO number densities as a function of mass at given values of the stream velocity. Here, unlike in the top panel, we integrate Equation (\ref{eq:NofMSIGO}) to infinity, rather than to the maximum mass we previously used to match simulations. We also find a simple relation between the SIGOs' cumulative number density as a function of mass $M_{\rm SIGO}$ and $\sigma_{\rm vbc}$ that fits the analytical model.
\begin{eqnarray}\label{eq:model}
    \log_{10} \left(  \frac{ N(>M_{\rm SIGO}) }{ {\rm Mpc}^3}\right) &= &  a(v_{\rm bc}) \bigg\{\log_{10} \left( \frac{M_{\rm SIGO}}{{\rm M}_\odot}\right) \bigg\}^{b(v_{\rm bc})}  \nonumber \\ &+&c(v_{\rm bc}) \ .
\end{eqnarray}
Fits for the prefactors $a(v_{\rm bc}), b(v_{\rm bc})$ and $c(v_{\rm bc})$ are given in Appendix \ref{Appendix-B}.

In the bottom left panel, we show the ratio of SIGO number densities at given stream velocities and masses to the mean SIGO number density in the Universe at their mass. In other words, we show:
\begin{equation}\label{eq:fSIGO}
    f_{\rm SIGO}=\frac{N(>M)_{\rm SIGO, vbc}}{\left<N(>M)_{\rm SIGO}\right>} \ ,
\end{equation}
with
\begin{equation}\label{eq:meanSIGO}
    \left<N(>M)\right>=\int^{\infty}_{0} p(v_{\rm bc}) N(>M)_{\rm SIGO, vbc}  dv_{\rm bc} \ ,
\end{equation}
where $p(v_{\rm bc})$ is the likelihood of a given stream velocity given by a Maxwell distribution, and $\left<N(>M)_{\rm SIGO}\right>$ is the mean number of SIGOs per comoving Mpc$^3$ at or above mass M in the Universe. Note, we could not use the number density of SIGOs with no stream velocity as the denominator for this ratio as in Figure \ref{fig:MtotComp}, because SIGOs are only found when the stream velocity effect is present. As shown in Figure \ref{fig:MSIGOComp}, left panel, large values of the stream velocity ($\sim 3.5 \sigma_{\rm vbc}$) result in an enhancement in SIGO abundances on the order of $20-30\times$ the mean abundance of SIGOs in the Universe. However, we note that only a small fraction of the Universe by volume (about $5\times10^{-8}$) have stream velocities above $3.5 \sigma_{\rm vbc}$. We can also see that larger stream velocity values result in larger maximum SIGO masses. 



Using our analytical approach, we can now examine the large-scale abundance of SIGOs. In particular, in Figure \ref{fig:SIGOMap} we depict the fluctuations in SIGO abundances resulting from the variations in stream velocity on large scales. Analogous to the DM case, we show  
\begin{equation}\label{eq:deltaSIGO}
    \delta_{\rm SIGO} = \frac{N(>M)_{\rm vbc} - \left<N(>M)\right>}{\left<N(>M)\right>} \ ,
\end{equation}
with $N(>M)$ as defined in Equation (\ref{eq:NofMSIGO}). The overdensities in the plot were generated using the method for generating appropriately distributed density fields outlined in Section \ref{ssec:Analytic}.


We also present a power spectrum of the fluctuations in SIGO abundances (i.e., fluctuations in $N[M_{\rm SIGO}>10^5 M_\odot]$) on large scales, the bottom right panel of Figure \ref{fig:CombinedPSPlot}. The power spectrum of these number densities is calculated as 
\begin{equation}\label{eq:DMPS}
    \left<N_{\mathbf{k},\rm SIGO}N^*_{\mathbf{k},\rm SIGO}\right> = (2\pi)^3 P(k) \delta^{(3)}(\mathbf{k}-\mathbf{k'}) \ .
\end{equation}
The plot shows the variance of these number density fluctuations per $\ln k$: $\Delta^2 (k) \equiv k^3 / (2\pi^2) \times P(k)$. Also shown is the power spectrum of M$ > 10^6 {\rm M_\odot}$ DM halo abundances, for comparison, both with the effects of large-scale density fluctuations (main figure) and with only the effects of the stream velocity (inset figure). Note the similarities between the inset figure and the power spectrum of SIGO abundances, which are primarily set by velocity fluctuations.
Note that the coherence scale of the number density is set by the range of scales over which $\Delta^2(k)$ is nonzero. In this case, as with the stream velocity that gives rise to this effect, that scale is approximately $k > 0.5$ Mpc$^{-1}$, implying that number densities are coherent on scales of a few comoving Mpc.


In the top panel of Figure \ref{fig:CombinedPSPlot}, we show the probability density of observing an abundance $N_{\rm SIGO}(M > 10^5$~M$_\odot)$ of SIGOs per Mpc$^3$ within a region of the Universe with constant (but unknown) stream velocity. This can be expressed mathematically as $p(N_{\rm SIGO}) = p(v_{\rm bc})\times dv_{\rm bc} / d(N_{\rm SIGO})$, where $p(v_{\rm bc})$ is given by a Maxwell distribution with scale parameter $\sigma_{\rm vbc} / \sqrt{3}$. To facilitate quick calculations for future semi-analytical studies, we provide a fit (also showed in the figure) for this probability density function: 
\begin{align}
&\begin{aligned}
\begin{split}\label{eq:pSIGOApprox}
{\rm log}_{10}[p(N_{\rm SIGO})] & \approx -0.933\times {\rm exp}[1.734 \times {\rm log}_{10}(N_{\rm SIGO})] \\ & +0.261\times {\rm exp}[-0.507\times {\rm log}_{10}(N_{\rm SIGO})],\\ 
\end{split}
\end{aligned}
\end{align}
calculated using a chi-squared fit with a Trust Region algorithm, where $N_{\rm SIGO}$ is given in units per Mpc$^3$ comoving.
 Note that this fit diverges as N$_{\rm SIGO}\rightarrow0$, and is valid for N$_{\rm SIGO}\gtrsim 0.001$ Mpc$^{-3}$.

Using this equation, we see a fairly high likelihood that a given (small) region of space will contain virtually no SIGOs ($\sim29\%$ chance that a given region will contain fewer than 0.1 SIGOs per Mpc$^3$), but a long tail--there is about a $13\%$ chance that a given region of space will contain more than 1 SIGO per Mpc$^3$.

If indeed SIGOs are the progenitors of globular clusters, we would expect the distribution of globular clusters on the sky, as well as their overall abundance, to be similar. Using our model, we can predict the average abundance of SIGOs in the local Universe at early redshifts. Note that here we have used $\sigma_8 = 1.7$, in order to enhance the abundance of SIGOs in our simulations. However, we may still use this figure to compare to the real Universe for several reasons. The density power spectrum scales as $P(k)\propto a^2 \approx z^{-2}$, so doubling the normalization of the power spectrum as we have is similar (at the redshifts and scales we are discussing) to increasing the redshift discussed by a factor of $\sqrt{2}$ \citep[][]{park+20}. In other words, we can expect comparable SIGO abundances in the real Universe at $z\sim14$, which is consistent with our analytical calculations. In addition, clusters in the Universe tend to form on high sigma peaks of the large-scale density field \citep[e.g.,][]{Kaiser1984,Sheth_1999,BarkanaLoeb2004,Topping_2018}. Because of our proximity to the Virgo cluster, it is probable that we are at such a high sigma peak, which will enhance the concentration of structures of all masses relative to this work, which assumed a density consistent with the average matter density of the Universe.

\begin{figure*}[t]
\includegraphics[width=1.\textwidth]{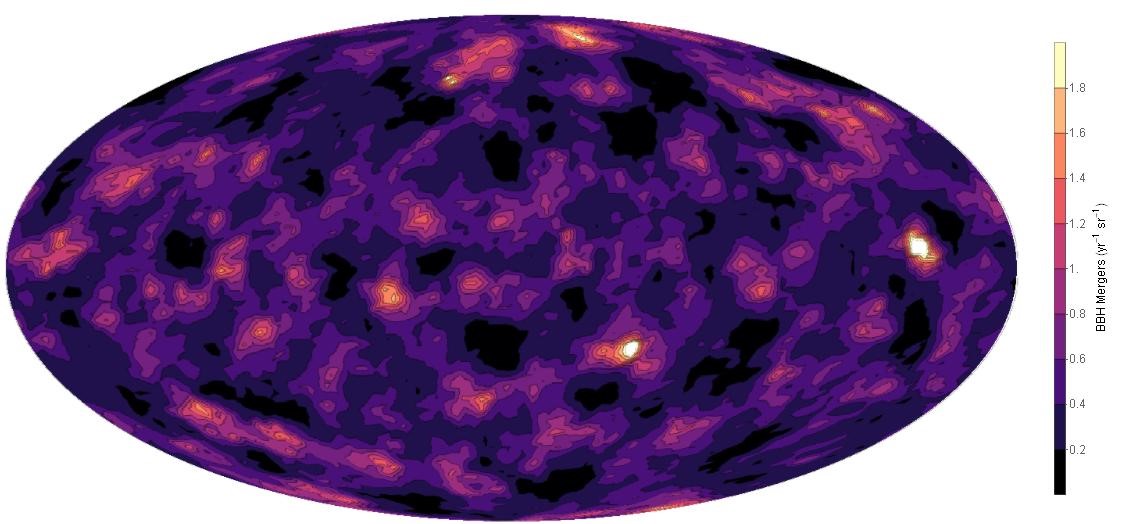}
\caption{Sky-map of integrated BBH merger abundances  in globular clusters, to a distance of 675 Mpc. Numbers are given in mergers per steradian per year, assuming a merger rate of $10^{-8}$ per year per cluster.}\label{fig:GWMap}
\end{figure*}

With that in mind, our model predicts an average SIGO number density of  $\sim 0.5$~Mpc$^{-3}$ above M $> 10^5$~M$_\odot$, which may be extrapolated to recent times, as was done in \citet{chiou+19}, yielding a possibly similar number density. This result is consistent to order of magnitude with the observed local density of globular clusters, estimates of which range from $0.72$ Mpc$^{-3}$ \citep[][]{Rodriguez+15} to a few $\times$ Mpc$^{-3}$ \citep[e.g.,][]{Zwart_2000,Harris+13}. Since SIGOs are early structures, they are likely to have low metallicities. \citet[]{Rodriguez+15} estimates the local density of low metallicity GCs as $0.46$ Mpc$^{-3}$, which is also consistent with our estimate for the abundance of SIGOs. It is worthwhile, however, to be cautious with these comparisons, as not all of these early SIGOs necessarily evolve into present-day globular clusters \citep[][]{naoznarayan14,popa,Chiou+21}, and as additional SIGOs are expected to form after the redshifts considered, increasing the overall number of SIGOs formed \citep[][]{naoznarayan14,popa,chiou18,chiou+19}. An additional consideration is the possibility that SIGOs will fragment in their collapse from their initial size, which is on the order of tens to hundreds of pc \citep[e.g.,][]{Chiou+21}, to the size of a globular cluster ($\sim 1$~pc). A typical SIGO, modelled as a puffy disc, has a Toomre stability criterion above unity \citep[][]{toomre}, and therefore will not fragment (excepting some central overdense regions, which could potentially collapse further to create high density star forming regions) in its collapse into GCs. In addition, in connecting the mass of SIGOs to the mass of globular clusters, we have implicitly assumed a high star formation efficiency. However, cluster formation and star formation are not 100\% efficient \citep[e.g.,][]{Baumgardt+07,Krumholz+19,Li+19,Grudic+20}, and while preliminary studies by \citet[]{Chiou+21} do support a relatively high star formation efficiency in SIGOs, star formation in SIGOs is still not fully understood. Nonetheless, these comparisons highlight an additional possible connection between these SIGOs and (particularly low-metallicity) globular clusters. 





\section{Implications to Gravitational Wave Anisotropies}\label{sec:GW}




For a standard initial stellar mass function, thousands of stellar-mass BHs likely form in a typical GC, many of which are likely initially retained \citep[]{Belczynski_2006, Willems2005, Wong2012}. On a timescale of $\lesssim 1\,$Gyr, these BHs sink to their host clusters' centers through dynamical friction, where they  dynamically interact with other BHs to form BH binaries \citep[e.g.,][]{SigurdssonHernquist1993}. It has been suggested that this process may be one of the leading sources for binary BH mergers \footnote{Although other processes are suggested to be comparable, from isolating binaries \citep[e.g.,][]{deMink+16,Belczynski+16,Marchant+16,Breivik+19,Breivik+20} to dynamical evolution at nuclear star clusters at the center of galaxies \citep[e.g.,][]{Hoang+18,Stephan+19,Wang+20}.} 
\citep[]{Zwart_2000,OLeary_2006,Rodriguez+15,Rodriguez+16,Chatterjee+17,Rodriguez+21}.

If indeed most BBH mergers form in GCs, and if SIGOs are indeed the main progenitor of GCs, there should be an anisotropy in GW signals from binary BHs derived from and comparable to the anisotropies in SIGO abundances due to spatial variations in stream velocity. Using our analytic model, we can estimate the expected observed anisotropy. 

In a typical GC similar to the ones observed in the Milky Way at present \citep[{mass of roughly a few$\times10^5\,M_{\odot}$, core radii of roughly $1\,$pc, metallicity of roughly $10\%$ solar; e.g.,}][]{Harris+96}, recent models predict roughly 100 total binary BH mergers over a roughly $10\,$Gyr cluster lifetime \citep[e.g.,][]{Rodriguez+18,FragioneKocsis18,Kremer+20,AntoniniGieles20}. Assuming that these mergers are roughly uniformly distributed in time at zero-th order, this implies a BBH merger rate of roughly $10^{-8}\,\rm{yr}^{-1}$ per GC. By combining this order-of-magnitude rate estimate with the expected spatial distribution of GCs linked to SIGOs, we build a sense of the potential anistropy in BBH mergers from GCs.

As a proof-of-concept, in Figure \ref{fig:GWMap} we show a sky-map of a line-of-sight integrated GW merger abundance, from {GW} BBH mergers in GCs, up to a distance of $675$~Mpc (determined by our analytical box, see Section \ref{ssec:Analytic}). At this distance redshift effects on the GW are negligible. As shown, we predict the integrated rate of BBH mergers may vary by as much as an order-of-magnitude over scales of $\sim 10\deg$. To a distance of $675$~Mpc, we also estimate an approximate rate of BBH mergers of $0.5$ mergers sr~$^{-1}$ yr$^{-1}$, with a standard deviation of approximately $0.3$ mergers sr$^{-1}$ yr$^{-1}$ across the sky.  \citet{Payne+20} found that ten black hole mergers from the first LIGO/Virgo catalog are consistent with an isotropic distribution, over large scales ($>100$~Mpc). Further, recent endeavors by \citet{LIGOSkyMap+21} to search for anisotropic stochastic GW backgrounds did not find anisotropies on the three directions in the sky. Thus, both analyses imply that larger scales than the ones predicted here do not exhibit anisotropies, yielding a possible clear GW-sky signatures of SIGOs.  As the catalog of binary BH mergers continues to grow through both current and ongoing LIGO/Virgo/KAGRA detections \citep[e.g.,][]{Abbott+20} and detections by proposed third-generation detectors such as the Einstein Telescope \citep[e.g.,][]{Punturo+10} and Cosmic Explorer \citep[e.g.,][]{Abbott+17_3G}, this anisotropy may potentially be observable and if observed, would further constrain the connection between GCs and SIGOs.

We stress, however, that Figure \ref{fig:GWMap} and this discussion represent an ideal case where we assume that \textit{all} SIGOs are directly linked to globular clusters and that their distribution (as well as the distribution of BH mergers) does not vary with redshift. While SIGOs may be linked to globular clusters \citep[e.g.,][]{naoznarayan14,chiou+19}, it is unlikely that all SIGOs become globular clusters \citep[e.g.,][]{popa,Chiou+21}, and we have yet to test the redshift evolution of SIGOs over large ranges. Nevertheless, this result suggests that if indeed SIGOs are GCs' progenitors, they may imprint an anisotropic sky distribution on the GW emission signal\footnote{Note that a \citet{Kroupa+01} initial mass function is often invoked; however, it is still unknown whether initial mass functions in the early Universe will follow a Kroupa profile. As was mentioned in \citet[]{Rodriguez+15}, a 1$\sigma$ variation in the slope of the high-mass end of the IMF can cause significant variation in the abundance of BBHs.} .

\section{Discussion}\label{sec:discussion}

Supersonically induced gas objects (SIGOs) containing little to no DM are expected to exist in the early Universe (before reionization) with masses of $\lsim {\rm few} \times 10^6$~M$_\odot$ \citep{naoznarayan14,popa,chiou18,chiou+19,Chiou+21}. They are the result of a decoupling between the DM and baryon fluids at the time of recombination because of the relative velocity between them \citep[a.k.a. stream velocity][]{TH}. This stream velocity is coherent only on small scales ($\sim$ few Mpc), which means that numerical simulations that track this effect can do so, over those scales. This small coherent scale poses a challenge when exploring large-scale SIGO abundances using numerical simulations. In addition, the relatively small mass of SIGOs ($\sim$few$\times 10^5-10^6$~M$_\odot$) also poses a mass-resolution challenge for numerical simulations.

Here, we combined small scale numerical simulations with analytical perturbation theory, and explored the large-scale distribution of SIGOs.  
Using normalizations obtained from high-resolution, small scale, numerical simulation results, we connect the decrease in dark matter halo formation at large stream velocities to SIGO abundances. We demonstrate that perturbation theory can be used to adequately model dark matter halo abundances, by comparing the results of perturbation theory to simulation results (Figure \ref{fig:MtotComp}). We additionally show a comparison between our model results and simulations of SIGO formation and abundance (Figure \ref{fig:MSIGOComp}), demonstrating that our model is useful for predicting SIGO abundances at typical stream velocities.

Our major results are as follows:

(i) \textbf{Halo abundance}: We show that increasing stream velocity decreases the number density of dark matter halos at $M <$ few~$\times 10^7 M_\odot$ in both our analytic model and simulations, consistent with previous studies \citep[e.g.,][]{TH, naoz12, popa} and with each other (Figure \ref{fig:MtotComp}). Using this model and a power spectrum of the distribution of stream velocities on the sky, we present simulated maps of the number density of dark matter halos at a given mass (Figure \ref{fig:DMMap}). For a comparison more directly related to our Universe, we also show a power spectrum of dark matter halo abundances across the sky (Figure \ref{fig:CombinedPSPlot}).

(ii) \textbf{SIGO abundance}: We use our analytic model for DM halo abundances to estimate SIGO abundances, giving the first fully analytic model for SIGO number densities. We posit that the decline in halos due to the stream velocity can be linked directly to the increase in SIGOs, and use this relation to analytically model SIGO number densities, using the abundance of SIGOs in simulations at various stream velocities to normalize our results (Figure \ref{fig:MSIGOComp}). The strong agreement between our analytical model and the small-box simulation results, as depicted in the top panel of Figure \ref{fig:MSIGOComp}, motivates us to use our analytical model to calculate the SIGOs abundance on large scales. 

(iii) \textbf{Anisotropy in the distribution of SIGOs}: Because prior simulations relied on constant stream velocities, they could not be extended to large scales. Our analytic model does not have this limitation, and can be therefore be used to measure the large-scale variations of SIGO quantities on the sky. We use this to simulate maps of SIGO distributions (Figure \ref{fig:DMMap}), and to create a power spectrum of SIGO abundances (Figure \ref{fig:CombinedPSPlot}). Our model predicts an average of $0.5$ SIGOs of mass $M > 10^5 M_\odot$ per comoving Mpc$^{3}$ at $z\sim14$, with a standard deviation $\sigma_{\rm SIGO} = 0.6$ Mpc$^{-3}$.

We also use the probability of a given stream velocity (given by a Maxwell distribution), along with the relation between stream velocity and SIGO number densities, to compute a probability density function for varying SIGO abundances (Equation (\ref{eq:pSIGOApprox}) and the top panel of Figure \ref{fig:CombinedPSPlot}).

(iv) {\bf Connection to high redshift observations:}
Simulations such as those in \citet[]{chiou+19} suggest that SIGOs occupy a distinctive region in luminosity-size parameter space that may be distinguishable in future JWST observations (specifically, SIGOs are predicted to be dimmer than classical objects of the same radius). Follow-up studies may therefore soon be able to place observational constraints on the abundance of SIGOs as well as on their variation across the sky, contributing additional physical insight to these results. Should JWST indeed be able to observe them, we would expect their large-scale abundances to vary based on a power spectrum in agreement with that presented in Figure \ref{fig:CombinedPSPlot}. The distribution of observed SIGOs should be qualitatively similar to the map presented in Figure \ref{fig:SIGOMap}.

(v) {\bf Connections to globular clusters:}

If SIGOs are a progenitor of globular clusters, we can connect our conclusions about the variation in SIGO abundances on the sky to GC abundances. We find a mean SIGO number density of $\sim 0.5$~Mpc$^{-3}$ at $z\sim14$, which is consistent to order of magnitude with the observed local density of globular clusters, estimates of which range from $0.72$ Mpc$^{-3}$ \citep[][]{Rodriguez+15} to a few $\times$ Mpc$^{-3}$ \citep[e.g.,][]{Zwart_2000,Harris+13}. More notably, this analytic SIGO number density is almost equal to the observed abundance of low metallicity GCs, $0.46$ Mpc$^{-3}$ \citep[][]{Rodriguez+15}. As SIGOs are early structures with correspondingly low (expected) metallicities, this could be another indicator of a connection between SIGOs and (particularly low-metallicity) GCs.

(vi) {\bf Anisotropy in the distribution of Gravitational Wave emission due to BBH merger events:} As discussed in Section \ref{sec:GW}, if we assume that SIGOs are indeed connected to GCs, we can connect the abundance of SIGOs to the abundance of GCs and therefore possibly to the abundance of BBH merger events. As a result, we suggest an anisotropy in the distribution of BBH merger events derived from the variation in SIGO abundances on the sky. As shown in Figure \ref{fig:GWMap}, this anisotropy could cause the integrated abundance of BBH mergers to vary by as much as an order of magnitude over scales of approximately $10\deg$, to a distance of 675 Mpc, in an idealized case (though we caution that that variation would decrease on longer sightlines). Future observations of BBH merger events from LIGO, Virgo, KAGRA, and others may be able to observe this anisotropy, and could therefore further constrain the relation between SIGOs and GCs.

To summarize, we presented the larger scale abundance of supersonically induced gas objects (SIGOs), using a combination of analytical and simulation approaches. We thus predict variation of these high density gas objects in the early Universe ($z\sim14$) with an average of $\sim 0.5$~Mpc$^{-3}$, possibly observable by JWST. The average number density of SIGOs is consistent with the local number density of  globular clusters, further supporting the proposal that these SIGOs are the progenitors of globular clusters \citep[e.g.,][]{naoznarayan14,chiou+19}. Finally, since globular clusters are natural birthplaces of black hole binary mergers, we propose that SIGOs may leave a distinct anisotropic signature on the gravitational wave signal on the sky.  



\acknowledgments
We thank the anonymous referee for useful comments. 
We also thank Naoki Yoshida and Carl Rodriguez for useful discussions. 
W.L., S.N., Y.S.C, B.B., F.M., and M.V. thank the support of NASA grant No. 80NSSC20K0500 and the XSEDE AST180056 allocation, as well as the
Simons Foundation Center for Computational Astrophysics and the UCLA cluster \textit{Hoffman2} for computational resources. F.M. acknowledges support from the Program ``Rita Levi Montalcini'' of the Italian MUR.  W.L. thanks Ryan Carlson for the helpful conversations. S.N thanks Howard and Astrid Preston for their generous support. Y.S.C thanks the partial support from UCLA dissertation year fellowship. B.B. also thanks the the Alfred P. Sloan Foundation and the Packard Foundation for support. MV acknowledges support through NASA ATP grants 16-ATP16-0167, 19-ATP19-0019, 19-ATP19-0020, 19-ATP19-0167, and NSF grants AST-1814053, AST-1814259,  AST-1909831 and AST-2007355. K.K. is supported by an NSF Astronomy and Astrophysics Postdoctoral Fellowship under award AST-2001751.

\appendix
\section{Procedure for normalizing analytic model}\label{sec:appendix}

There are two steps in producing the normalization factor $A$. First, we match the analytical results to each simulation run at each given value of $v_{\rm bc}$ using Equation (\ref{eq:NSIGOs}). For completeness, we show the equation again here, with terms labelled as coming from simulations or analytic models: 
\begin{equation}\label{eq:NSIGOs2}
    \frac{dN(u_{\rm vbc})}{dM}\bigg |_{\rm SIGO, Sim} = A(u_{\rm vbc})  \left( \frac{dN(u_{\rm vbc}=0)}{dM}\bigg |_{\rm DM, Analytic}-\frac{dN(u_{\rm vbc})}{dM}\bigg |_{\rm DM, Analytic} \right) \ 
\end{equation}

where we remind the reader that $u_{\rm vbc} = v_{\rm bc}/\sigma_{\rm vbc}$ and where the normalization factor is assumed to take the form:
\begin{equation}
 A(u_{\rm vbc}) = C \times  u_{\rm vbc}^{n} \ ,
\end{equation}

\begin{figure}[t]
\begin{centering}
\includegraphics[width=0.5\textwidth]{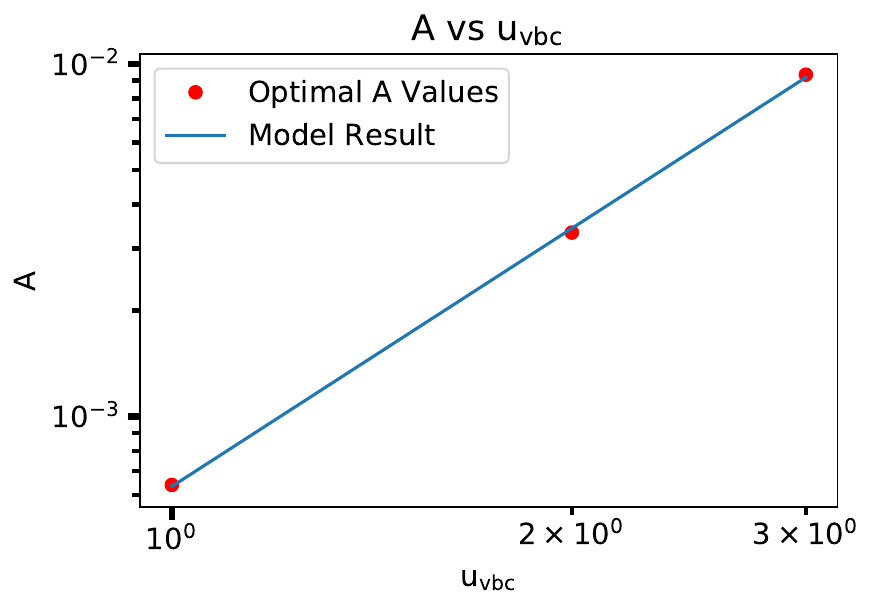}
\caption{Plot of the agreement between our model normalization parameter and our optimal normalization parameters obtained from comparison with simulations, as a function of stream velocity.}\label{fig:FitPlot}
\end{centering}
\end{figure}

 We use the Levenburg-Marquardt method to find the optimal value of $A$ for each simulated stream velocity: $u_{\rm vbc}=1$, $2$, and $3$. We limited this fit to masses above $4\times10^4 M_\odot$, as below this mass, the effects of our limited resolution affect our simulation data. These normalization values and their corresponding stream velocities are depicted in Figure \ref{fig:FitPlot} as red points. They are also listed in the table below, with error estimates $\delta A$: 

\begin{table}[bp]
\centering
\begin{tabular}{l l l l}
\hline
$u_{\rm vbc}$ & $1$                   & $2$                   & $3$                   \\ \hline
A                         & $6.40 \times 10^{-4}$ & $3.33 \times 10^{-3}$ & $9.35 \times 10^{-3}$ \\ \hline
${\rm \delta A}$          & $1.45 \times 10^{-5}$ & $6.12 \times10^{-5}$  & $6.65 \times 10^{-5}$ \\ \hline
\end{tabular}
\caption{Optimal values of our normalization factor $A$ with error estimates for each of our simulated stream velocities.}
\label{Table:App}
\end{table}

Second, we fit the points for all three simulations for the different $v_{\rm bc}$ values, using a linear model fit in log-log space. We find best-fit values of $C$ and $n$, using the above data for $A$, we find that $C = 6.3 \times 10^{-4} \pm 2.1\times 10^{-5}$ and $n = 2.43 \pm 0.04$, shown in Figure \ref{fig:FitPlot} as the blue line.

\section{Parameters for approximate SIGO number density model}\label{Appendix-B}

In order to enable future semi-analytic studies of SIGO number densities, we compute an approximate formula to estimate SIGO abundances as a function of stream velocity and mass. This formula provides results corresponding to the output of Equation (\ref{eq:NofMSIGO}). 

\begin{figure}[t]
\begin{centering}
\includegraphics[width=0.6\textwidth]{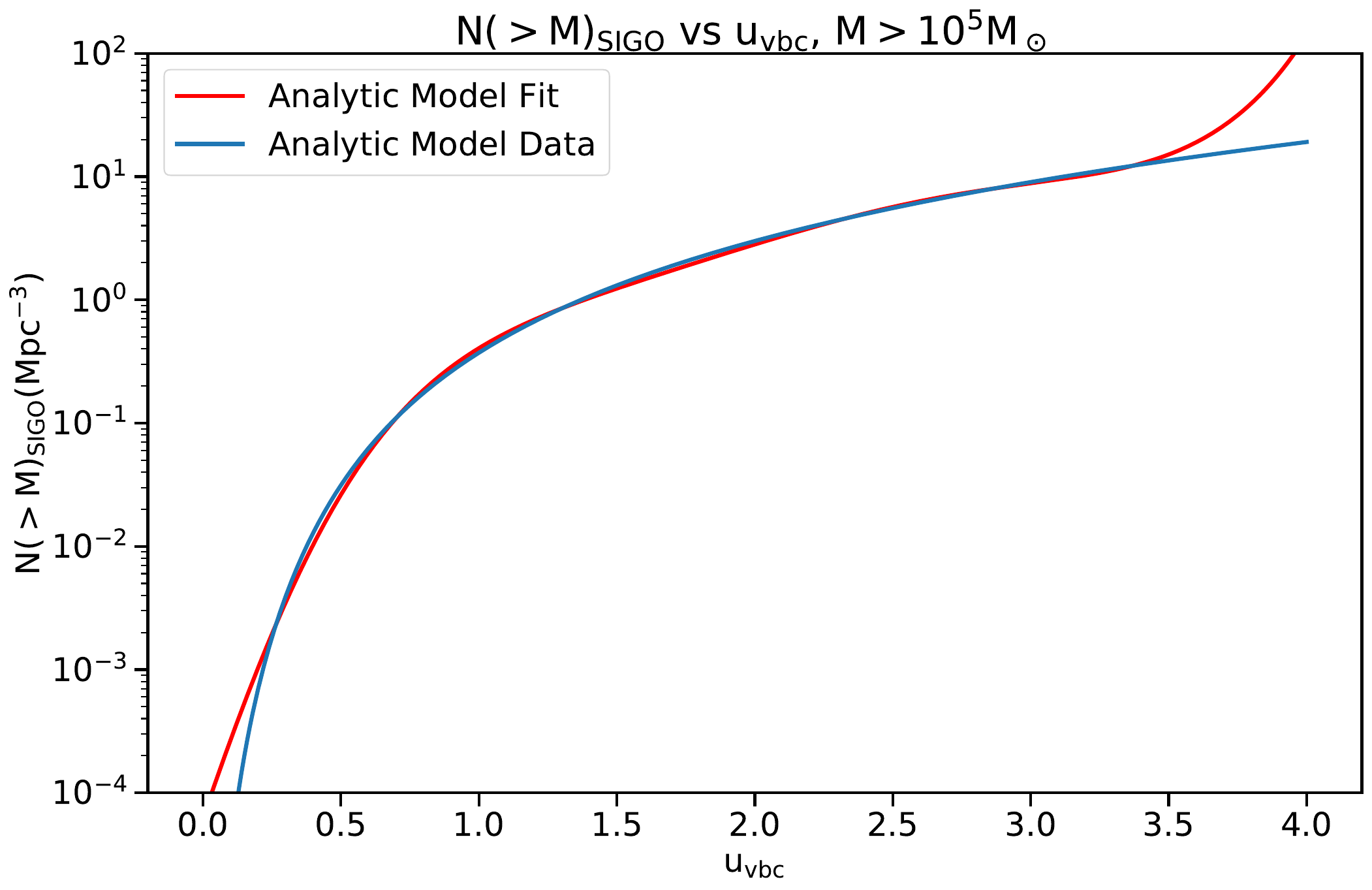}
\caption{Plot illustrating the agreement between our analytic model for SIGO abundances (Equation (\ref{eq:NofMSIGO})) and our reported fit to the model output (Equation (\ref{eq:model})) at M$ > 10^5$~M$_\odot$. The two equations agree closely for few~$\times 10^4$~M$_\odot <$~M~$<8\times 10^5$~M$_\odot$, and for $u_{\rm vbc} < 3.4$.}\label{fig:FitPlotNofM}
\end{centering}
\end{figure}

We found a form for this relation given by Equation (\ref{eq:model}), repeated here for convenience:
\begin{eqnarray}\label{eq:modelApp}
    \log_{10} \left(  \frac{ N(>M_{\rm SIGO}) }{ {\rm Mpc}^3}\right) &= &  a(u_{\rm vbc}) \bigg\{\log_{10} \left( \frac{M_{\rm SIGO}}{{\rm M}_\odot}\right) \bigg\}^{b(u_{\rm vbc})}  \nonumber + c(u_{\rm vbc}) \ ,
\end{eqnarray}
where
\begin{eqnarray}\label{eq:factors}
   a(u_{\rm vbc}) & =&  a_1 u_{\rm vbc}^5 + a_2 u_{\rm vbc}^4 + a_3 u_{\rm vbc}^3 + a_4 u_{\rm vbc}^2 + a_5 u_{\rm vbc} + a_6 \ ,\\
   b(u_{\rm vbc}) & =&  b_1 u_{\rm vbc}^5 + b_2 u_{\rm vbc}^4 + b_3 u_{\rm vbc}^3 + b_4 u_{\rm vbc}^2 + b_5 u_{\rm vbc} + b_6 \ ,\\
   {\rm and} && \nonumber \\
   c(u_{\rm vbc}) &= & c_1 u_{\rm vbc}^5 + c_2 u_{\rm vbc}^4 + c_3 u_{\rm vbc}^3 + c_4 u_{\rm vbc}^2 + c_5 u_{\rm vbc} + c_6 \ .
\end{eqnarray}
Using a nonlinear least squares regression, we find the best fit parameters to match this model to our analytic results. The best fit parameters are reported in Table \ref{tab:fit-parameters}. An example of the agreement between our analytic model and the fit to the model is presented in Figure \ref{fig:FitPlotNofM}. The results of Equation (\ref{eq:NofMSIGO}) are shown in blue, giving N$(>$M$)_{\rm SIGO}$ for M $>10^5 $M$_\odot$, with the result of Equation (\ref{eq:model}) shown in red. Notably, this reported fit holds for $2.5 \times 10^4$~M$_\odot \lesssim$~M~$\lesssim 8\times 10^5$~M$_\odot$, and for $v_{\rm bc} < 3.4 \sigma_{\rm vbc}$.

\begin{table}[htp]
\begin{minipage}{.9 \textwidth}\centering
\begin{tabular}{l l l l l l l}
\cline{1-7}
                        & $1$       & $2$      & $3$       & $4$        & $5$       & $6$        \\ \hline
\multicolumn{1}{l}{a} & $0.0014$  & $-0.0165$ & $0.0745$  & $-0.159$   & $-0.1554$ & $-0.1066$ \\ \hline
\multicolumn{1}{l}{b} & $0.0104$  & $-0.1248$ & $0.5661$  & $-1.186$ & $1.051$ & $2.067$   \\ \hline
\multicolumn{1}{l}{c} & $0.0718$ & $-0.7767$ & $3.239$ & $-6.618$   & $7.284$ & $-1.238$   \\ \hline
\end{tabular}
\end{minipage}
\caption{Best fit parameters used in Equation (\ref{eq:model}) that match analytic model results from Equation (\ref{eq:NofMSIGO}), obtained using a nonlinear least squares regression. Coefficients are given to high precision due to the sensitivity of the model to the coefficients.}
\label{tab:fit-parameters}
\end{table}

\bibliography{myBib}{}
\bibliographystyle{aasjournal}



\end{document}